\documentclass{elsarticle}
\pdfoutput=1
\usepackage{etex}
\usepackage{graphicx} 
\usepackage{hyperref} 
\usepackage{tabularx} 
\usepackage{mdframed} 
\usepackage{xspace} 
\usepackage{color}
\usepackage{verbatim}
\usepackage{paralist}
\usepackage{verbatim} 
\usepackage{listings}
\usepackage{multirow}
\usepackage{algorithm} 
\usepackage{algorithmic}

\usepackage{framed}

\newcommand{\todo}[1]{\textcolor{red}{TODO: #1}}
\newcommand{\inred}[1]{\textcolor{red}{#1}}

\newcommand{\bold}[1]{{{\textbf{#1}}} \xspace}

\newcolumntype{L}[1]{>{\hsize=#1\hsize\raggedright\arraybackslash}X}
\newcolumntype{R}[1]{>{\hsize=#1\hsize\raggedleft\arraybackslash}X}
\newcolumntype{C}[2]{>{\hsize=#1\hsize\columncolor{#2}\centering\arraybackslash}X}
\newcolumntype{s}{>{\hsize=.5\hsize}X}
\newcolumntype{b}{>{\hsize=1.16\hsize}X}
\newcommand{\headtable}[1]{{\bf{#1}} \xspace}
\newcommand*\rot{\rotatebox{90}} 

\definecolor{codegray}{gray}{0.9}

\newcommand{\mycode}[1]{\lstinline[basicstyle=\small]$#1$}

\newcommand{\epfl}{EP\_FL}
\newcommand{\epmpgranularity}{EP\_MPG}
\newcommand{\epmpselection}{EP\_MPS}
\newcommand{\epmoperatordef}{EP\_OD}
\newcommand{\epmoperatorselection}{EP\_OS}
\newcommand{\epnavstrat}{EP\_NS}
\newcommand{\eprefiningpatch}{EP\_SP}
\newcommand{\eppatchvalidation}{EP\_PV}
\newcommand{\epfitnessfunction}{EP\_FF}
\newcommand{\epingredientpool}{EP\_IPD}
\newcommand{\epingredientselection}{EP\_IS}
\newcommand{\epingredienttransformation}{EP\_IT}
\newcommand{\uniquerepaired}{11}

\begin{document}

\title{Astor: Exploring the Design Space of Generate-and-Validate Program Repair beyond GenProg}

\author[add1]{Matias Martinez}\author[add2]{Martin Monperrus}

\address[add1]{Universit\'e Polytechnique Hauts-de-France, France}
\address[add2]{KTH Royal Institute of Technology, Sweden}

\begin{abstract}
This article contributes to defining the design space of program repair. Repair approaches can be loosely characterized according to the main design philosophy, in particular ``generate- and-validate'' and synthesis-based approaches. Each of those repair approaches is a point in the design space of program repair. 
Our goal is to facilitate the design, development and evaluation of repair approaches by providing a framework that: a) contains components commonly present in most approaches, b) provides built-in implementations of existing repair approaches.
This paper presents a Java framework named Astor that focuses on the design space of generate-and-validate repair approaches. The key novelty of Astor is to provides explicit extension points to explore the design space of program repair. Thanks to those extension points, researchers can both reuse existing program repair components and implement new ones. Astor includes 6 unique implementations of repair approaches in Java, including GenProg for Java called jGenProg.
Researchers have already defined new approaches over Astor. The implementations of program repair approaches built already available in Astor are capable of repairing, in total, 98 real bugs from 5 large Java programs.
Astor code is publicly available on Github: \url{https://github.com/SpoonLabs/astor}.

\end{abstract} 

\maketitle

\section{Introduction}

Automated software repair is a research field that has emerged during the last decade for repairing real bugs of software application. The main goal is to reduce cost and time of software maintenance by proposing to developers automatically synthesized patches that solve bugs present in their applications.
Among pioneer repair systems are GenProg \cite{Weimer2009},  Semfix \cite{Nguyen:2013:SPR}, Prophet \cite{prophet}, Nopol \cite{nopol}, and others \cite{Kim2013,Xiong2017,Mechtaev2016,Le2017JSR,Le2017SSS,Ke2015RPS,prophet,Long2017AIC,Perkins2009,SANER2017,Durieux2016DDC,weimer2013AE,directfix,qi2014strength,spr}.
Automation of bug fixing is possible by using \emph{automated correctness oracles}.
For instance, GenProg \cite{Weimer2009} introduced the use of \emph{test suite} as correctness oracle: the correctness of a bug fix is is assessed by executing all tests from its associated test suite.

Program repair systems can be loosely characterized along their main design philosophy: generate-and-validate approaches (which first generate a set of candidate patches, and then validate them against the test suite) or synthesis based approaches (which first use test execution information to build a repair constraint, and then use a constraint solver to synthesize a patch).
For example, GenProg and JAFF \cite{ArcuriEvolutionary} are generate-and-validate approaches based on \emph{genetic programming}.

More generally, every repair system is a point in the design space of program repair.
By making design decisions explicit in that design space, one can start to have a fine-grain understanding of the core conceptual differences in the field.
For example, the main conceptual difference between GenProg and PAR \cite{Kim2013} lies in the repair operators: they do not use the same code transformations for synthesizing patches.

To foster research on program repair, we aim at providing the research community with a generic framework that encodes the code design space of generate-and-validate repair approaches.

In this paper, our main contribution is Astor (\underline{A}utomatic \underline{S}oftware \underline{T}rans-formations f\underline{O}r program \underline{R}epair). Astor is a program repair framework for Java, it provides 6 \emph{generate and validate} repair approaches: jGenProg (a Java implementation of GenProg, originally in C), jKali (an implementation of Kali \cite{Qi2015}, originally in C), jMutRepair (an implementation of MutRepair \cite{debroy2010using}, not publicly available), DeepRepair (\cite{ white2017dl}), Cardumen \cite{CardumenArxiv}, and  TIBRA (an extension of jGenProg introduced in this paper).
Those repair approaches are based on twelve extension points that form the first ever explicit design space of program repair. Over those twelve extension points, the program repair researchers can  both choose an existing component (among 33 ones), or implement new ones for exploring a a new point in the design space of program repair.

\begin{framed}
Astor has been extensively used by the research community 
\begin{enumerate}[\it a)]
\item for creating a novel repair system based on Astor \cite{Tanikado2017NewStrategies,white2017dl,Zu2017Test4Repair,gpfl2017}, which is the key enabling factor, 
\item for performing comparative evaluations, using Astor's publicly available implementation of existing approaches \cite{xin2017leveraging},
\item for reusing numerical results and/or patches obtained with Astor \cite{Xiong2017,le2016history,Saha:2017:EEO,Arja1712.07804}.
\end{enumerate}
\end{framed}

Astor is publicly available on Github and is actively maintained. A user community is able to provide support. Bug fixes and extensions are welcome as external contributions (pull requests).
From an open-science perspective, since the whole code base is public, peer researchers can validate the correctness of the implementation and hence minimize threats to internal validity.

To sum up, our contributions are:
\begin{itemize}
\item The explicit design space of generate-and-validate program repair.
\item The realization of that design space in Astor, where the most important design decisions of program repair are encoded as extension point.
\item Twelve extension points, which program repair researchers can  either reuse or extend for doing  interesting research in the field.
\item Six repair approaches that can be used out-of-the-box in comparative evaluations, incl. jGenProg, the most used Java implementation of GenProg according to citation impact {\cite{Weimer2009,LeGoues2012TSEGP,LeGoues2012,Weimer2010,forrest2009genetic}}. The study of the repair capability of the implemented approaches based on the Defects4J bug benchmark \cite{JustJE2014}.
\item The evaluation of two extension points of GenProg: choice of the ingredient space and ingredient transformation.
\end{itemize}

This paper is a completely rewritten long version of a short paper \cite{astor2016}. It includes a detailed explanation of Astor's architecture, extension pointss as well as a large evaluation.
The paper continues as follows.
Section \ref{sec:architecture} describes the design of Astor, 
Section \ref{sec:extensionpoints} presents the extension points provided by Astor.
Section \ref{sec:approachessummary} presents the built-in approaches included in Astor.
Section \ref{sec:evaluation} presents a evaluation of the built-in approaches and different implementations for the extension points.
Section \ref{sec:relatedwork} presents the related work.
Finally, section \ref{sec:conclusion} concludes the paper.

\section{Architecture}
\label{sec:architecture}

\subsection{The Design of Astor}
\label{sec:framework}

Astor is a framework that allows researchers to implement new automated program repair approaches, and to extend  available repair approaches such as jGenProg \cite{defects4j-repair} (implementation of GenProg \cite{Weimer2009}), jKali \cite{astor2016} (implementation of Kali \cite{Qi:2015:APP:2771783.2771791}), jMutRepair  \cite{astor2016} (implementation from MutRepair \cite{debroy2010using}), DeepRepair \cite{white2017dl}, Cardumen \cite{CardumenArxiv}.

Astor encodes the design space of \emph{generate-and-validate} repair approaches, which first search within a search space to generate a set of patches, and then validate using a correctness oracle. 
Astor provides twelve extension points that form the design space of generate-and-validate program repair.
New approaches can be implemented by choosing an existing component for each extension point, or to implement new ones.

The extension points allow Astor users to define the design of a repair approach.
Main design decisions are:
\begin{inparaenum}[\it a)]
\item  code transformations (aka repair operators) used to define the solution search space;
\item  different strategies for navigating the search space of candidate solutions; 
and 
\item mechanism for validating a candidate solution.
\end{inparaenum}

Astor was originally conceived for building \emph{test-suite based repair approaches} \cite{Weimer2009} and the first implemented approach over it was named jGenProg, a Java implementation of GenProg \cite{Weimer2009}, originally written in OCaml language for repairing C code.
In test-suite based repair, test suites are considered as a proxy to the program specification, and a program is considered as fulfilling its specification if its test suite passes all the these cases otherwise, the program has a defect. The test suite is used as a \emph{bug oracle}, i.e., it asserts the presence of the bug, and as \emph{correctness oracle}.

An approach over Astor requires as input a buggy program to be repaired and a correctness oracle such as a test suite.
As output, the approach generates, when it is possible, one or more patches that are valid according to the correctness oracle.

\begin{algorithm}[t]
\begin{algorithmic}[1]
\REQUIRE{Program under repair $P$ }
\REQUIRE{test suite $TS$ }
\ENSURE{A list of test-suite adequate patches}

    \STATE{suspicious $\leftarrow$ run-fault-localization(P, TS)} \label{algomain:fl} //\epfl
     \STATE{mpl $\leftarrow$ create-modification-points(suspicious)} \label{algomain:mp} //\epmpgranularity
  \STATE{ops  $\leftarrow$ get-operators()} //\epmoperatordef \label{algomain:retrieveopts}
    \STATE{tsa-patches-refined $\leftarrow \emptyset$}
      \STATE{ nr-iteration $\leftarrow$ 0} \label{algomain:iterationinit}
       \STATE{starting-time $\leftarrow$ System.currentTime}

     \WHILE{continue-searching(starting-time, nr-iteration,  size(tsa-patches))} \label{algomain:evaluatenav} 
      \STATE{program-variants $\leftarrow$ generate-program-variants(P, mpl, ops)} \label{algomain:generate}
       \STATE{tsa-patches $\leftarrow$ tsa-patches + validate-variants(P, program-variants, TS)}   \label{algomain:validate}
      \label{algomain:navigation}
        \STATE{nr-iteration $\leftarrow$ nr-iteration + 1} \label{algomain:iterationincrease}
     \ENDWHILE //\epnavstrat \label{algomain:endnavigation}
   \STATE{tsa-patches-refined $\leftarrow$  refining-patches(tsa-patches)}   \label{algomain:refining} //\eprefiningpatch
\RETURN{tsa-patches-refined}  \label{algomain:return}
\end{algorithmic}
\caption{Main steps of \emph{generate-and-validate} repair approaches, implemented in Astor (extension points are referred to as comment prefixed by \texttt{//}, )}
\label{alg:main}
\end{algorithm}

Algorithm \ref{alg:main} displays  the high-level steps executed, in sequence, by a \emph{generate-and-validate} repair approach built in Astor.
They are: 
\begin{inparaenum} [\it 1)]
\item Fault localization (line \ref{algomain:fl}), 
\item Creation of a representation (line \ref{algomain:mp}),
\item Navigation of the search space (lines \ref{algomain:evaluatenav}-\ref{algomain:endnavigation}),  and
\item Solution post-processing (line \ref{algomain:refining}).
\end{inparaenum}
In the remainder of this section, we describe each step.

\subsection{Fault Localization}
\label{sec:fault-localization}

The \emph{fault localization} step is the first step executed by an approach built over Astor. It aims at determining what is wrong in the program received as input.
This step is executed on line \ref{algomain:fl} of Algorithm \ref{alg:main}.
Fault localization consists of computing locations that are suspicious. 
In the context of repair, fault localization allows to reduce the search space by discarding those code locations that are probably healthy. 
A repair approach can use the suspiciousness values of locations to guide the search in the solution space. Consequently, fault localization has an impact on the effectiveness of the repair approach \cite{mao2012FL}.

Test-suite based repair approaches from Astor use fault localization techniques based on spectrum analysis. Those techniques execute the test cases of a buggy program and trace the execution of software components (e.g., methods, lines). Then, from the collected traces and the tests results (i.e., fail or pass), the techniques use formulas to calculate the suspicious value of each component. The suspicious value goes for 0 (lowest probability that the component contains a bug) to 1 (highest). 
Repair approaches use different formulas, for instance, GenProg uses an ad-hoc formula  \cite{Weimer2009}, while MutRepair \cite{debroy2010using} uses the Tarantula formula \cite{Jones2002}.

Astor provides fault localization as an extension point named \epfl, where researchers can plug implementations of any fault localization technique.
Astor provides a component (used by default) that implement that point and uses the fault localization library named GZoltar  \cite{gzoltar2012} and the Ochiai formula \cite{abreu2006evaluation}.\footnote{http://www.gzoltar.com/}

\subsection{Identification of Modification Points}

Once the fault localization step returns a list of suspicious code locations (such as statements), 
Astor create a representation of the program under repair.

\begin{framed}
{\underline{Definition 1}}: A \emph{Modification point} is  a code element (e.g., a statement, an expression) from the buggy program under repair that can be modified with the goal of repairing the bug.
\end{framed}

Astor creates modification points from the  suspicious statements returned by the fault localization step (section \ref{sec:fault-localization}). 
This step is executed in line \ref{algomain:mp} of Algorithm \ref{alg:main}.
Astor provides an extension point named \epmpgranularity{} (section \ref{sec:exgranularity}) to define the granularity of each modification point according to  that one targeted by a repair approach built over Astor. 
For instance, jGenProg creates one modification point per each statement indicated as suspicious by the fault localization. 
Cardumen, another approach built over Astor, works at a fine-grained level: it creates a modification point for each expression contained in a suspicious statement. 
Other approaches focus on particular code elements, such as  jMutRepair which  creates modification points only for  expressions with unary and binary operators.
For example, let us imagine that the fault localization marks as suspicious the two lines presented in Listing \ref{listing1}.
\begin{lstlisting}[caption={Two suspicious statements},basicstyle=\scriptsize, label={listing1}]
 9      ....
10	myAccount = getAccount(name); 
11	myAccount.setBalance(previousMonth + currentMonth);	
12  	....
\end{lstlisting}

jGenProg creates two modification points, both pointing to statements, one to the assignment at line 10, another to the method invocation at line 11.
Contrary, Cardumen creates 3 modifications points: one pointing to the expression at the right size of the assignment at line 10, a second one to the method invocation at line 11 (note that the method invocation it is also an expression), and the last one pointing to the expression (\mycode{previousMonth + currentMonth}) which is the parameter of the method invocation at line 11.

\subsection{Creation of repair operators}
\label{sec:operatorspacedef}

Astor synthesizes patches by applying automated code transformation over suspicious modification points. 
Those transformations are done by \emph{repair operators} and the set of all repair operators that an approach considers during the repair conform the \emph{repair operator space}.

\begin{framed}
\underline{Definition 2}: a \emph{Repair operator} is an action that transforms a code element  associated to a modification point into another, modified compatible code element.
\end{framed}

Astor provides an extension point named \epmoperatordef{} (section \ref{sec:eposdef}) for specifying the operator space that a repair approach will use.
The extension point is invoked at line \ref{algomain:retrieveopts} of Algorithm \ref{alg:main}.
Astor works with two kinds of repair operators:

\paragraph{Synthesis and repair operators}
An approach can synthesize new code by directly applying one transformation operators to a modification point, without the need of any extra information (in particular without using ingredients).
One of them is the repair operator from jMutRepair which changes a logical operator from \mycode{>} to \mycode{>=}. For example, it generates the new code \mycode{(fa * fb) >= 0.0} from the code \mycode{(fa * fb) > 0.0} without the use of any further information.

\paragraph{Synthesis based on ingredients}
\label{sec:ingredients-based}

There are operators that need some extra information before applying a code transformation in a modification point $mp_i$.
For instance, two operators (Insert and Replace) from GenProg \cite{Weimer2009} need one statement (aka the \emph{ingredient}) taken from somewhere in the application under repair. 
Once selected, the ingredient is inserted before or replace the code at $mp_i$.
Such approaches are known as \emph{Ingredient-based repair approaches} \cite{Weimer2009}.
jGenProg  and DeepRepair \cite{white2017dl} are two ingredient-based approaches built over Astor.
Astor gives support to such repair approaches by automatically providing:
\begin{inparaenum}[\it a)]
\item a pool of ingredients available, 
\item strategies for selecting ingredients from the pool.
\end{inparaenum}

\subsection{Navigation of the search space}
\label{sec:navss}

Once that that all code transformations to be applied to a suspicious element are known,
Astor proceeds with the navigation of the search space. 
The goal is to find, between all possible modified versions of the buggy program, one or more versions that do not contain the bug under repair and that do not introduce new bugs.

The navigation of the search space is executed by the main loop (Algorithm \ref{alg:main} line \ref{algomain:evaluatenav}). 
In each iteration, the approach verifies whether a set of code changes done by repair operators over some modification points (which produce a modified version of the program) repair the bug.
A \emph{modified} version of a buggy program is called in Astor a \emph{Program variant}.

\begin{framed}
\underline{Definition 3}: a \emph{Program variant} 
is a entity that stores:
\begin{inparaenum}[\it a)]
\item one or more code locations known as modification point;
\item the repair operators applied to each  modification point;
\item the code source resulting from the execution of all repair operators over the corresponding modification points.
\end{inparaenum}
\end{framed}

Then, Astor computes a candidate \emph{patch} from a program variant.
A \emph{Patch} produced by Astor is a set of changes between the buggy version and a modified version represented by a program variant.

Astor allows to override the navigation strategy using the extension point \epnavstrat{}  (section \ref{sec:epnavstr}).
Algorithm \ref{alg:main} presents the default implementation of the navigation strategy, named \emph{Selective}, which executes the two main steps that characterize a \emph{generate-and-validate} technique:
first, the algorithm generates 1+ program variants (Line \ref{algomain:generate}), then it validates them against the test suite (Line \ref{algomain:validate}).

\subsubsection{Generation of Program Variants}

\begin{algorithm}[t]
\begin{algorithmic}[1]
\REQUIRE{Program under repair $P$ }
\REQUIRE{List of modification points $MPs$}
\REQUIRE{List of operator $OS$ }
     \STATE{mps $\leftarrow$ choose-modification-points(MPs)} //\epmpselection
     \label{algoexplo:chooseMP}
      \STATE{transformations $\leftarrow \emptyset$}
      \FORALL{mp-i $\in$  mps }
        \STATE{op-j$\leftarrow$ choose-operator(mp-i, OS)}  \label{algoexplo:chooseop} //\epmoperatorselection
         \STATE{transformations $\leftarrow$ transformations $\cup$ create-transformation(P, mp-i, op-j)} \label{algoexplo:addt}
         
        \label{algoexplo:createpatches} 
     \ENDFOR
      \STATE{program-variants $\leftarrow$ apply-transformations(transformations, P)} \label{algoexplo:applyt} 
\RETURN{program-variants}  \label{algoexplo:return}
\end{algorithmic}
\caption{Method \emph{generate-program-variants} (as comment prefixed by \texttt{//}, the name of the extension point)}
\label{alg:generation}
\end{algorithm}

Algorithm \ref{alg:generation} shows the main steps that Astor executes for creating a program variant. Let us analyze each of them.

\paragraph{Selection of modification points}
First, a repair approach built over Astor chooses, according to a selection strategy, the modifications points (at least one) to apply repair operators (Algorithm \ref{alg:generation} line \ref{algoexplo:chooseMP}).
Astor provides an extension point named \epmpselection{} (section \ref{sec:epnavigationsusp}) for specifying customized strategies of modification points selection.

\paragraph{Selection of repair operators}

For each selected modification point $mp_i$, an approach selects one repair operator $op_j$ to apply at  $mp_i$ (Algorithm \ref{alg:generation} line \ref{algoexplo:chooseop})
and adds the transformation created by \emph{create-transformation} (Algorithm \ref{alg:createtransformation}) to the set of code transformations (Algorithm \ref{alg:generation}  line \ref{algoexplo:addt}).
Astor provides an extension point named \epmoperatorselection{} (section \ref{sec:epopselection}) for specifying customized strategies of operator selection.

\begin{algorithm}[t]
\begin{algorithmic}[1]
\REQUIRE{Program under repair $P$ }
\REQUIRE{Modification point $MP$}
\REQUIRE{Repair operator $OP$ }

\IF{needs-ingredient($OP$)}\label{algotrans:needsingredient} 
\IF{pool-not-initialized(ingredient-pool)}
 \STATE{ingredient-pool $\leftarrow$ build-pool(P)} //\epingredientpool \label{algotrans:ip} 
\ENDIF
 \STATE{ingredient $\leftarrow$  select-ingredient(ingredient-pool)} // \epingredientselection \label{algotrans:is} 
 \STATE{transformed-ingredient $\leftarrow$ transform-ingredient(ingredient)} //\epingredienttransformation  \label{algotrans:it} 
\RETURN{($MP$, $OP$, transformed-ingredient)} \label{algotrans:returningredient} 
\ELSE
\RETURN{($MP$, $OP$)} \label{algotrans:returnnormal} 
\ENDIF
\end{algorithmic}
\caption{Method \emph{create-transformation} (as comment prefixed by \texttt{//}, the name of the extension point)}
\label{alg:createtransformation}
\end{algorithm}

\paragraph{Creation of code transformation}

In Astor, a code transformation is a concept that groups a modification point $mp$ and a repair operator $op$. Algorithm \ref{alg:createtransformation} shows the creation of code transformations.
When the operator is of kind ingredient-based  (section \ref{sec:ingredients-based}) the transformation needs an additional element: an  ingredient for synthesizing a patch.
Astor provides to those operators a pool that contains all the ingredients that they can use.

For creating a transformation, Astor first detects if the operator needs ingredients or not (Algorithm \ref{alg:createtransformation} line \ref{algotrans:needsingredient}).
If it does not need any ingredient, Astor returns the transformation composed by the $mp$ and $op$ (Algorithm \ref{alg:createtransformation} line \ref{algotrans:returnnormal}).
Otherwise,  
Astor  creates an \emph{ingredient pool} from the program under repair.
The creation involves to first parse the code at a given granularity (by default, that one given by the extension point \epmpgranularity{}) and  then store each parsed element in the pool (Algorithm \ref{alg:createtransformation} line \ref{algotrans:ip}). Astor provides an extension point named \epingredientpool{} (section \ref{sec:epingspacedef}) for plugging a customized strategy for building the ingredient pool.

The ingredient pool is queried by the repair approach when an ingredient-based  repair operator needs an ingredient for synthesizing the candidate patch code (Algorithm \ref{alg:createtransformation} line \ref{algotrans:is}).
Astor provides an extension point named \epingredientselection{} (section \ref{sec:epingselect}) for plugging in a customized strategy in order to select an ingredient from the ingredient pool.

Moreover, when an operator gets an ingredient for the ingredient pool, it can use it directly  (i.e., without applying any transformation) or after applying a transformation over the ingredient (Algorithm \ref{alg:createtransformation} line \ref{algotrans:it}).
For instance, a transformation proposed by Astor is to replace variables from the ingredient that is not in the scope of the modification point. Astor provides an extension point named \epingredienttransformation{} (section \ref{sec:epingtransf}) for plugging a customized strategy of ingredients transformation.

\paragraph{Creation of program variants}
\label{sec:synthesis}

A repair approach over Astor generates program variants from the code transformation previously generated  (Algorithm \ref{alg:generation} line \ref{algoexplo:applyt}) and returns them for  validation stage (line \ref{algoexplo:return}).

\subsubsection{Candidate patch validation}
\label{sec:validation}

Algorithm \ref{alg:validation} shows the main steps that Astor executes for validating program variants and returning test-suite adequate patches. 
Once program variants are created, 
the Astor framework synthesizes from each variant the code source of the patch (Algorithm \ref{alg:validation} line \ref{algoval:patchsynt}), then applies it to the buggy version of the program under repair and finally evaluates the modified version (line \ref{algoval:verifypatch}) using the correctness oracle.
If the patched version is valid, the corresponding patch is a solution and it is stored (line \ref{algoval:solution}).

Astor provides an extension point named \eppatchvalidation{} (section \ref{sec:epvalidation}) for specifying the validation process to be used by the repair approach.
Built-in repair approaches over Astor use test-suite as specification of the program \cite{Weimer2009} and as correctness oracle. No failing test cases means the program is correct according to the specification encoded on the test suite. To validate candidate patches, Astor runs the test suite on the patched version of the buggy program.

Moreover, Astor defines an extension point named \epfitnessfunction{} (section \ref{sec:epff}) to specify the \emph{Fitness Function} that evaluates the patch using the output from the validation process (Algorithm \ref{alg:validation} line \ref{algoval:ff}).
The result of this function is used to determine if a patch is a solution (i.e., repair the bug) or not.
By default, the fitness function on Astor counts the number of failing test cases.
No failing test case means the patch is a solution and is known as \emph{test-suite adequate patch}.

\begin{algorithm}[t]
\begin{algorithmic}[1]
\REQUIRE{Program under repair $P$ }
\REQUIRE{List of program $program-variants$}
\REQUIRE{Test suite $TS$ }
    \FORALL{pv-i $\in$ program-variants}
     \STATE{patch-i $\leftarrow$ synthesize-patch-from-variant(P, pv-i)} \label{algoval:patchsynt}
    \STATE{validation-result $\leftarrow$ validate(TS, P, patch-i, pv-i)} //\eppatchvalidation
    \label{algoval:verifypatch}
    \IF{is-valid(validation-result) //\epfitnessfunction} 
       \label{algoval:ff}
            \STATE{tsa-patches $\leftarrow$ tsa-patches $\cup$ pc-i}  \label{algoval:solution} 
        \ENDIF
    \ENDFOR

\RETURN{tsa-patches}  // test suite adequate patches  \label{algoval:return}
\end{algorithmic}
\caption{Method \emph{validate-variants} (as comment prefixed by \texttt{//}, the name of the extension point)}
\label{alg:validation}
\end{algorithm}

\subsection{Evaluation of conditions for ending navigation}

An approach over Astor finishes the search of patches, i.e., stops the loop at line \ref{algomain:evaluatenav} from Algorithm \ref{alg:main},   when any of these conditions is fulfilled (configured by the user):
\begin{inparaenum}[\it a)]
\item finding $n$ plausible patches, 
\item iterating $n$ times, 
\item executing during $h$ hours (timeout).
\end{inparaenum}

\subsection{Solution post-processing}

After finishing navigating the search space, Astor provides an extension point named \eprefiningpatch{} (section \ref{sec:leppriority}) for processing the patches found, if any.
We envision  two kinds of  post-processing.
First, the post-processing of each patch found aims at applying, for instance, patch minimization or code formatting.
As proposed by the GenProg \cite{Weimer2009}, some changes done in a solution program variant could not be related to the bug fixing. A post-processing aims at removing such changes and keeping only those that are necessary to repair the bug.
Second, the post-processing of the list of patches aims sorting patches according to a given criterion.
By default, Astor lists the patches found in chronological order (first patch found, first patch listed). However, as the number of patches could be large, a repair system could order patches according to, for instance, their location, to the number of modifications each introduce, type of modification, etc.

\newcommand{\titleepvalues}{Implemented components}

\section{Extension points provided by Astor}
\label{sec:extensionpoints}

\newcolumntype{b}{>{\hsize=0.55\hsize}X} 
\newcolumntype{s}{>{\hsize=.25\hsize}X}
\newcolumntype{f}{>{\hsize=.2\hsize}X}

\begin{table}[p]
\def\arraystretch{1.5}
\centering
\tiny
\begin{tabularx}{\textwidth}{|s||s|b|}
\hline
 \headtable{Extension point} &  \headtable{Component} &  \headtable{Explanation} \\ 
\hline
\hline
\headtable{Fault}
  & GZoltar& Use of third-party library GZoltar\\ \cline{2-3}
\headtable{localization (\epfl)} & CoCoSpoon& Use of third-party library CoCoSpoon\\ \cline{2-3}
 \hline

 \headtable{Granularity}
 
 \headtable{modification}
 
  \headtable{points (\epmpgranularity)}
&Statements & Each modification point corresponds to a \bold{statement} and repair operators generate code at the level of \emph{statements}\\ \cline{2-3}

&Expression & Each modification point corresponds to an expression and repair operators generate code at the level of \emph{expressions} \\ \cline{2-3}

 &logical-relational-operators & Modification points target to binary expression whose operators are  logical (AND, OR) or relational (e.g., > ==) \\ \cline{2-3}
  &if conditions & Modification points target to the expression inside \emph{if} conditions\\ \cline{2-3}
 \hline
 
\headtable{Navigation}
  &Exhaustive & Complete navigation of the search space \\ \cline{2-3}
 \headtable{strategy (\epnavstrat)}
 &Selective & Partial navigation of search space guided, by default, by random steps\\ \cline{2-3}
  &Evolutionary & Navigation of the search space using genetic algorithm\\ \cline{2-3}
 \hline
 
 \headtable{Selection of}
 
 \headtable{suspicious}
  &Uniform-random & Every modification point has the same probability to be changed by an operator\\ \cline{2-3}
\headtable{modification}

 \headtable{points (\epmpselection)}
  &Weighted-random & The probability of changed of a modification point depends on the suspiciousness of the pointed code\\ \cline{2-3}

  &Sequential & Modification points are changes according to the suspiciousness value, in decreasing order \\ \cline{2-3}
 \hline

 \headtable{Operator space}  
 
  \headtable{definition (\epmoperatordef)} 
 &IRR-statements & Insertion, Removement and Replacement of statements\\ \cline{2-3}
 &Relational-Logical-op& Change of unary operators, and logical and relational binary operators\\ \cline{2-3}
 &Suppression& Suppression of statement, Change of if conditions by True or False value, insertion of remove statement\\ \cline{2-3}
  &R-expression& replacement of expression by another expression\\ \cline{2-3}
 \hline
 
 \headtable{Selection of}  
 
 \headtable{operator (\epmoperatorselection)} 
 &Uniform-Random & Every repair operator has the same probability of be chosen to modify a modification point  \\ \cline{2-3}

 &Weighted-Random & Selection of operator based on non-uniform probability distribution over the repair operators. \\ \cline{2-3}
 \hline
 
\headtable{Ingredient pool}  

\headtable{definition (\epingredientpool)} 
&File & Pool with ingredients written in the same file where the patch is applied.\\ \cline{2-3}
 
&Package & Pool with ingredients written in the same package where the patch is applied.\\ \cline{2-3}
 
&Global & Pool with all ingredients from the application under repair. \\ \cline{2-3}
 \hline
 \headtable{Selection of} 
 &Uniform-random & Ingredient randomly chosen from the ingredient pool\\ \cline{2-3}
  \headtable{ingredients (\epingredientselection)} 
 
  &Code-similarity-based & Ingredient chosen from similar method to that where the candidate patch is written\\ \cline{2-3}
   &Name-probability-based & Ingredient chosen based on the frequency of its variable's names \\ \cline{2-3}
 \hline
 \headtable{Ingredient}  
 &No-Transformation & Ingredients are not transformed\\ \cline{2-3}
 \headtable{transformation (\epingredienttransformation)}  
  &Random-variable-replacement & Out-of-scope variables from an ingredients are replaced by randomly chosen in-scope variables\\ \cline{2-3}
   &Name-cluster-based & Out-of-scope variables from an ingredients are replaced by similar named in-scope variables\\ \cline{2-3}
   &Name-probability-based & Out-of-scope variables from an ingredients are replaced by in-scope variable based on the frequency of variable's names \\
 \hline
  \headtable{Candidate patch} 
  
   \headtable{Validation (\eppatchvalidation)}
 &Test-suite & Original test-suite used for validating a candidate patch \\ \cline{2-3}
  
  &Augmented-test-suite & New test cases are generated for augmented the original test suite used for validation\\ \cline{2-3}
 \hline
  \headtable{Fitness function for}
  
   \headtable{evaluation (\epfitnessfunction)}
 &Number-failing-tests & The fitness is the number of failing test cases. Lower is better. Zero means the patch is a test-suite adequate patch\\ \cline{2-3}

 \hline
 \headtable{Solution} 
 
 \headtable{prioritization (\eprefiningpatch)} 
 &Chronological& Generated valid patches are printing Chronological order, according with the time they were discovered\\ \cline{2-3}

 &Less-Regression &Patches are presented according to the number of failing cases from those generated test cases, in ascending order\\ \cline{2-3}
 \hline
\end{tabularx}
\label{tab:extension_points_values}
\caption{Summary of extension points and components already implemented in Astor.}
\end{table}
 
In this sectionn we detail the main extension points that are provided by the Astor framework for creating new repair approaches.
For each extension point we give the name and description of the components already included in the framework.
Table \ref{tab:extension_points_values} summarizes all extension points.

\bigskip
\subsection{Fault Localization (\epfl)}
\label{sec:exfloc}

\subsubsection{\titleepvalues}
\begin{itemize}
\item GZoltar: use of third-party library GZoltar.
\item CoCoSpoon: use of third-party library CoCoSpoon.
\item {\emph{Custom}: name of class that implements interface \mycode{FaultLocalizationStrategy}}.
\end{itemize}

This extension point allows to specify the fault localization algorithm that Astor  executes (at Algorithm \ref{alg:main} line \ref{algomain:fl}) to obtain the buggy suspicious locations as explained in section \ref{sec:fault-localization}.
The extension point takes as input the program under repair and the test suite, and produces as output a list of program locations, each one with a suspicious value. 
The suspicious value associated to location $l$  goes between 0 (very low probability that $l$ is buggy) and 1  (very high probability that $l$ is buggy).
New fault localization techniques such that PRFL presented by Zhang et al. \cite{Zhang2017BSF} can be implemented in this extension point.

\bigskip
\subsection{Granularity of Modification points (\epmpgranularity)}
\label{sec:exgranularity}

\subsubsection{\titleepvalues}
\begin{itemize}
\item Statements: each modification point corresponds to a \emph{statement}. Repair operators generate code at the level of \emph{statements}.
\item Expressions: each modification point corresponds to an \emph{expression}. Repair operators generate code at the level of \emph{expressions}.
\item Logical-relational-operators: Modification points target to binary expression whose operators are  logical (AND, OR) or relational (e.g., $>,==$).
\item {\emph{Custom}: name of class that implements interface \mycode{TargetElementProcessor}}.
\end{itemize}

\subsubsection{Description}

The extension point \epmpgranularity{}  allows to specify the \emph{granularity}  of code that is manipulated by a repair approach over Astor.
The granularity impacts two components of Astor.
First, it impacts the program representation: Astor creates modifications points only for suspicious code elements of a given granularity (Algorithm \ref{alg:main} line \ref{algomain:mp}).
Second, it impacts the repair operator space: a repair operator takes as input code of a given granularity and generates a modified version of that piece of code. 
For example, the approach jGenProg manipulates \emph{statements}, i.e., the modification points refer to statements  and it has 3 repair operators: add, remove and replace  of statements.  
Differently, jMutRepair manipulates binary and unary expressions using repair operators that change binary and unary operators.

\bigskip

\subsection{Navigation Strategy (\epnavstrat)}
\label{sec:epnavstr}

\subsubsection{\titleepvalues}

\begin{itemize}
\item Exhaustive: complete navigation of the search space.
\item Selective:  partial navigation of search space guided, by default, by random steps.
\item Evolutionary: navigation of the search space using genetic algorithm.
\item {\emph{Custom}: name of class that extends class \mycode{AstorCoreEngine}}.
\end{itemize}

\subsubsection{Description}
The extension point \epnavstrat{} allows to define a strategy for navigating the search space.
Algorithm \ref{alg:generation} from section \ref{sec:navss} displays a \emph{general} navigation strategy, where 
most of its steps are calls to other extension points.
Astor provides three navigation strategies: \emph{exhaustive}, \emph{selective} and  \emph{evolutionary}.

\paragraph{Exhaustive navigation}
\label{sec:exhaustive}

This strategy \emph{exhaustively} navigates the search space, that is, all the candidate solutions are considered and validated.
An approach that carries out an exhaustive search  visits \emph{every} modification point $mp_i$  and applies to it \emph{every} repair operator $op_j$ from the repair operator space.
For each combination $mp_i$ and $op_j$, the approach generates zero or more candidates patches.
Then, for each synthesized $patch_i$  the approach applies it into the program under repair $P$ and then executes the validation  process as explained in section \ref{sec:validation}.

\paragraph{Selective navigation} 
\label{sec:guided_nav}
The selective navigation visits a portion of the search space.
This strategy is necessary when the search space is too large to be exhaustively navigated.
On each step of the navigation, it uses two strategies for determining
where to modify (i.e., modification points) and how (i.e., repair operators).
By default, the selective navigation uses weighted random for selecting modification points, where the weight is the suspiciousness value, and uniform random for selecting operators.
Those strategies can be customized using extension points \epmpselection{} (section \ref{sec:epnavigationsusp}) and \epmoperatorselection{} (section \ref{sec:epopselection}), respectively.

\paragraph{Evolutionary navigation}
\label{sec:evolution_nav}

Astor framework also provides the  Genetic Programming \cite{Koza:1992:GPP} technique for navigating the solution search space.
This technique was introduced in the domain of the automatic program repair by JAFF \cite{ArcuriEvolutionary} and GenProg \cite{Weimer2009}.
The idea is to evolve a buggy program by applying repair operators to arrive to a modified version that does not have the bug.
In Astor, it is implemented as follows: one considers an initial population of size $S$ of program variants and one evolves them across $n$ generations.
On each generation $i$, Astor first creates, for each program variant $pv$,  an offspring $pvo$ (i.e., a new program variant) and  applies, with a given probability,  repair operators to one or more modification points from $pvo$.
Then, it applies, with a given probability, the crossover operator between two program variants which involves to exchange one or more modification points. 
Astor finally evaluates each variant (i.e., the patch synthesized from the different operators applied) and then chooses the $S$ variants with best fitness values (section \ref{sec:validation}) to be part of the next generation.

\bigskip

\subsection{Selection of suspicious modification points (\epmpselection)}
\label{sec:epnavigationsusp}

\subsubsection{\titleepvalues}

\begin{itemize}
\item Uniform-random: every modification point has the same probability to be selected and later changed by an operator.
\item Weighted-random: the probability of changed of a modification point depends on the suspiciousness of the pointed code.
\item Sequential: modification points are changes according to the suspiciousness value, in decreasing order.
\item {\emph{Custom}: name of class that extends class \mycode{SuspiciousNavigationStrategy}}.
\end{itemize}

The extension point \epmpselection{} allows to specify the strategy to navigate the search space of suspicious components represented by modification points.
This extension point is invoked in every iteration of the navigation loop (Algorithm \ref{alg:main} line \ref{algomain:evaluatenav}): the strategy selects the  modification points where the repair algorithm will apply repair operators (Algorithm \ref{alg:generation} line \ref{algoexplo:chooseMP}).
Under the \emph{uniform random} strategy, each modification point $mp_x$ has the same probability of being selected, that is $p_u() = 1 / |MP| $, where $|MP|$ is the total number of modification points considered.
With the \emph{weighted random} strategy, each modification point has a particular probability of being selected computed as follows:
 $p_w(mp_x) =  \frac{sv_{mp_x}}{\sum_{j=1}^{|MP|} sv_{mp_j}}$, where $|MP|$ is the total number of modification points and $sv_{mp_x}$ is the  suspiciousness value of the modification point given by the fault localization algorithm.

\bigskip
\subsection{Operator spaces definition (\epmoperatordef)}
\label{sec:eposdef}

\subsubsection{\titleepvalues}

\begin{itemize}
\item IRR-Statements: insertion, removement and replacement of statements.
\item Relational-Logical-operators: change of unary operators, and logical and relational binary operators.
\item Suppression: suppression of statement, change of if conditions by True or False value, insertion of remove statement.
\item R-expression: replacement of expression by another expression.
\item {\emph{Custom}: name of class that extends class \mycode{OperatorSpace}}.
\end{itemize}

\subsubsection{Description}

After a modification point is selected, Astor selects a repair operator from the \emph{repair operator space} to apply into that point.
Astor provides the extension point \epmoperatordef{} for specifying the repair operator space  used by  a repair approach built on Astor. The extension point is invoked at line \ref{algomain:retrieveopts} of Algorithm \ref{alg:main}.
The operators space configuration depends on the repair strategy. 
For example, jGenProg has 3 operators (insert, replace and remove statement) whereas Cardumen has one (replace expression).

\bigskip

\subsection{Selection of repair operator (\epmoperatorselection)}
\label{sec:epopselection}

\subsubsection{\titleepvalues}

\begin{itemize}
\item Uniform-random: every repair operator has the same probability of being chosen to modify a modification point.
\item Weighted-random: selection of operator based on non-uniform probability distribution over the repair operators. Each repair operator has a particular probability of being chosen to modify a modification point.
\item {\emph{Custom}: name of class that extends \mycode{OperatorSelectionStrategy}}.

\end{itemize}

\subsubsection{Description}

The extension point \epmoperatorselection{} allows Astor's users to specify a strategy to select, given a modification point $mp$,  one operator from the operator space. 
By default, Astor provides a strategy that applies uniform random selection and it does not depend on the selected $mp$.
This strategy is used by approaches that uses selective navigation of the search space such as jGenProg and it is executed at line \ref{algoexplo:chooseop} from Algorithm \ref{alg:generation}. 
This extension point is useful  for implementing strategies based on probabilistic models such those presented by \cite{Martinez2013}.
In that work, several repair models are defined from different sets of bug fix commits, where each model is composed by repair operators and their associated probabilities calculated based on changes found in the commits. 

\bigskip

\subsection{Ingredient pool definition (\epingredientpool)}
\label{sec:epingspacedef}
\subsubsection{\titleepvalues}

\begin{itemize}
\item File: pool with ingredients written in the same file where the patch is applied.
\item Package: pool with ingredients written in the same package where the patch is applied.
\item Global: pool with all ingredients from the application under repair.
\item {\emph{Custom}: name of class that extends the class \mycode{AstorIngredientSpace}}.
\end{itemize}

\subsubsection{Description}

The ingredient pool contains all pieces of code that an ingredient-based repair approach can use for synthesizing a patch (section \ref{sec:operatorspacedef}).
For example, in jGenProg the ingredient pool contains all the statements from the application under repair. Then, jGenProg replaces a buggy statement by another one selected from the pool.
jGenProg can also add a statement before the suspicious one.
 
The extension point \epingredientpool{} allows to customize the creation of the ingredient pool. 
Astor provides three methods for building an ingredient pool, called ``scope'': file, package and global scope.
When ``file'' scope is used, the ingredient pool contains only ingredients that are in the same file where the patch will be applied (i.e., $mp$).
When the scope is ``package", the ingredient pool is formed with all the code from the package that contains the modification point $mp$. 
Finally, when the scope is ``global", the ingredient pool has all code from the program under repair. By default the original GenProg has a global ingredient scope, because it yields the biggest search space.
The ``file''  ingredient pool is smaller than the package-one, which is itself smaller than the global one.

\bigskip

\subsection{Selection of ingredients (\epingredientselection)} 
\label{sec:epingselect}

\subsubsection{\titleepvalues}

\begin{itemize}
\item Uniform-random: ingredient randomly chosen from ingredient pool.
\item Code-similarity-based: ingredient chosen from similar methods to the buggy method.
\item Name-probability-based: ingredient chosen based on the frequency of its variable's names.
\item {\emph{Custom}: name of class that extends class IngredientSearchStrategy}.
\end{itemize}

\subsubsection{Description}

The extension point \epingredientselection{} allows to specify the strategy that an ingredient-based repair approach from Astor uses for selecting an ingredient from the ingredient pool.
Between the implementations of this point provided by Astor,
One, used by default by jGenProg, executes uniform random selection for selecting an ingredient from a pool built given a scope (see section \ref{sec:epingspacedef}). Another, defined for DeepRepair approach, prioritizes ingredients that come from methods which are similar to the buggy method.  

\bigskip

\subsection{Ingredient transformation (\epingredienttransformation)}
\label{sec:epingtransf}

\subsubsection{\titleepvalues}

\begin{itemize}
\item No-transformation: ingredients are not transformed.
\item Random-variable-replacement: out-of-scope variables from an ingredients are replaced by randomly chosen in-scope variables.
\item Name-cluster-based: out-of-scope variables from an ingredients are replaced by similar named in-scope variables.
\item Name-probability-based: out-of-scope variables from an ingredients are replaced by in-scope variable based on the frequency of variable's names.
\item {\emph{Custom}: name of class that extends class \mycode{IngredientTransformationStrategy}}.
\end{itemize}

\subsubsection{Description}

The extension point \epingredienttransformation{} allows to specify the strategy used for transforming ingredients selected from the pool.
Astor provides four implementations of this extension point.
For instance, the strategy defined for DeepRepair approach replaces each out-of-scope variable form the ingredient by one variable in the scope of the modification points. The selection of that variable is based on a cluster of variable names, which each cluster variable having semantically related names \cite{white2017dl}.
Cardumen uses a probabilistic model for selecting the most frequent variables names to be used in the patch.
On the contrary, jGenProg, as also the original GenProg, does not transform any ingredient.

\bigskip

\subsection{Candidate Patch Validation (\eppatchvalidation)}
\label{sec:epvalidation}

\subsubsection{\titleepvalues}

\begin{itemize}
\item Test-suite: original test-suite used for validating a candidate patch.
\item Augmented-test-suite: new test cases are generated for augmented the original test suite used for validation.
\item {\emph{Custom}: name of class that extends class \mycode{ProgramVariantValidator}}.

\end{itemize}

\subsubsection{Description}

The extension point \eppatchvalidation{} executes the validation process of a patch (section \ref{sec:validation}).
Astor framework provides to test-suite based repair approaches a validation process that runs the test-suite on the patched program. 
The validation is executed in Algorithm \ref{alg:generation} line \ref{algoval:verifypatch}.

Another strategy implemented in Astor was called  MinImpact \cite{Zu2017Test4Repair},  proposed to alleviate the problem of patch overfitting \cite{smith2015cure}.
MinImpact uses additional automatically generated test cases to further check the correctness of a list of generated test-suite adequate patches and returns the one with the highest probability of being correct.
MinImpact implements the extension point \eppatchvalidation{}  by generating new test cases (i.e., inputs and outputs) over the buggy suspicious files, using Evosuite \cite{evosuite} as test-suite generation tool.
Once generated the new test cases, MinImpact executes them over the patched version.
The intuition is that the more additional test cases fail on a tentatively patched program, the more likely the corresponding patch is an overfitting patch.
MinImpact then sorts the generated patches by prioritizing those with less failures over the new tests.

Moreover, this extension point  can be used to measure other functional and not functional properties beyond the verification of the program correctness.
For example, instead of focusing on automated software repair, an approach built over Astor could target on minimizing the energy computation. For that, that approach would extend this extension point for measuring the consumption of a program variant.

\bigskip

\subsection{Fitness Function for evaluating candidate (\epfitnessfunction)}
\label{sec:epff}

\subsubsection{\titleepvalues}

\begin{itemize}
\item Number-failing-tests: the fitness is the number of failing test cases. Lower is better. Zero means the patch is a test-suite adequate patch.
\item {\emph{Custom}: name of class that implements \mycode{FitnessFunction}}.
\end{itemize}

\subsubsection{Description}

The extension point  \epfitnessfunction{} allows to specify the \emph{fitness function}, which consumes the output from the validation process of a program variant $pv$ and assigns to $pv$ its fitness value.
Astor provides an implementation of this extension point which considers as  fitness value  the number of failing test cases (low  is better).

On evolutionary approaches (section \ref{sec:evolution_nav}) such as jGenProg, this fitness function guides the evolution of a population of program variants throughout a number of generations.
In a given generation $t$, those variant with better fitness will be part of the population at generation $t+1$. 
On the contrary, on selective or exhaustive approaches, the fitness function is only used to determined if a patched program is solution or not.

\bigskip

\subsection{Solution prioritization (\eprefiningpatch)}
\label{sec:leppriority}

\subsubsection{\titleepvalues}

\begin{itemize}
\item Chronological: generated valid patches are printing chronological order, according with the time they were discovered.
\item Less-regression: patches are presented according to the number of failing cases from those generated test cases, in ascending order.
\item {\emph{Custom}: name of class that implements \mycode{SolutionVariantSortCriterion}}
\end{itemize}

\subsubsection{Description}

The extension point \eprefiningpatch{} allows to specify a method for sorting the discovered valid patches. 
By default, approaches over Astor  present patches sorted by time of discovery in the search space.
Astor proposes an implementation of this point named \emph{Less-regression}.
The strategy, defined by MinImpact \cite{Zu2017Test4Repair}, sorts the \emph{original test-suite} adequate patches with the goal of minimizing the introduction of regression faults, i.e., the approach prioritizes the patches with \emph{less failing test cases} from those tests automatically generated during the validation process.

\section{Repair Approaches implemented in Astor}
\label{sec:approachessummary}

\begin{table}[!t]
\def\arraystretch{1.5}
\centering
\tiny

\begin{tabularx}{\textwidth}{|l|X||X|X|X|X|X|X|}
\hline

\multicolumn{2}{|c||}{\headtable{Name Extension}} & \headtable{jGenProg}  & \headtable{jKali} & \headtable{jMutRepair} & \headtable{DeepRepair} & \headtable{Cardumen} & \headtable{TIBRA}\\ 
  \multicolumn{2}{|c||}{\headtable{Point}}  & \ref{sec:approachjgenprog}& \ref{sec:approachjkali} & \ref{sec:approachjmutation} & \ref{sec:approachdeeprepair} & \ref{sec:approachcardumen} &\ref{sec:approachtibra}\\ 
\hline

\hline
\ref{sec:exfloc} & \headtable{Fault localization (\epfl)} & GZoltar  & - & GZoltar & GZoltar & GZoltar & GZoltar\\ \hline
 \ref{sec:exgranularity} &  \headtable{Granularity of modification points (\epmpgranularity)}  & Statement & Statements + \emph{if} & Relational-Logical-operators& Statement & Expression & Statement \\ \hline
 \ref{sec:epnavstr} &  \headtable{Navigation strategy (\epnavstrat)}&  Selective or Evolutionary  & Exhaustive & Exhaustive or Selective  &Selective or Evolutionary  & Selective & Selective\\ \hline
 \ref{sec:epnavigationsusp} & \headtable{Selection susp. points (\epmpselection)} & Weighted-random& Sequence & Weighted-random& Weighted-random & Weighted-random & Weighted-random \\ \hline
 \ref{sec:eposdef} &  \headtable{Operator space definition (\epmoperatordef)} &  IRR-statements & Suppression & Relational-Logical-operators & Suppression & R-expression&Weighted-random\\ \hline
 \ref{sec:epopselection} &  \headtable{Selection operator (\epmoperatorselection)} &  Random & Sequential & Random & Random & -  & Random\\ \hline
 \ref{sec:epingspacedef} &  \headtable{Ingredient pool definition (\epingredientpool)} &  Package& - & - & Package & Global&Package\\ \hline
 \ref{sec:epingselect} &  \headtable{Selection ingredients (\epingredientselection)} &  Uniform-random & - & - & Code-similarity-based &Uniform-random & Uniform-random\\ \hline
 \ref{sec:epingtransf} &  \headtable{Ingredient transformation (\epingredienttransformation)} & No-transf. & - & - & Name-cluster-based & Name-probability-based &Random-variable-replacement\\ \hline
 \ref{sec:epvalidation} &  \headtable{Candidate patch validation (\eppatchvalidation)} & Test-suite &  Test-suite& Test-suite & Test-suite& Test-suite& Test-suite\\ \hline
 \ref{sec:epff} &  \headtable{Fitness function (\epfitnessfunction)} & \#failing-tests  & \#failing-tests & \#failing-tests & \#failing-tests & \#failing-tests& \#failing-tests\\ \hline
 \ref{sec:leppriority} &  \headtable{Solution prioritization (\eprefiningpatch)} & Chronol.  & Chronol. & Chronol. & Chronol. & Chronol. & Chronol. \\ \hline

\end{tabularx}
\label{tab:extension_points}
\caption{Main Extension points and decision adopted by each approach. Each approach and extension point includes a reference to the section that explain it.}
\end{table}

In this section we present a brief description of repair approaches built over Astor framework and publicly available at Github platform.
Those approaches were built combining  different components implemented for the extension points presented in section \ref{sec:extensionpoints}.
Table \ref{tab:extension_points} displays the components that form each built-in repair approach from Astor.
The approaches are presented in the order they were introduced into Astor framework.

\subsection{jGenprog}
\label{sec:approachjgenprog}

jGenProg is an implementation of GenProg \cite{Weimer2009} built over Astor framework.
The approach belongs to the family of ingredient-based repair approaches (section \ref{sec:ingredients-based}) and it has 3 repairs operators: insert, replace and  remove statements.
For the two first mentioned operators, jGenProg uses statements written somewhere in the application under repair for synthesizing patches that insert or replace statement.
jGenProg can navigate the search space (section \ref{sec:evolution_nav}) in two ways: 
\begin{inparaenum}[\it a)]
\item 
using evolutionary  search, as the original GenProg does; or
\item using  selective, as RSRepair \cite{qi2014strength} does.
\end{inparaenum}

\subsection{jKali}
\label{sec:approachjkali}

The technique Kali was presented by \cite{Qi:2015:APP:2771783.2771791} for evaluating the incompleteness test suites used by repair approaches for validating candidate patches. The intuition of the authors was that removing code from a buggy application was sufficient  to pass all test from incomplete test suite. Consequently, the generated patches overfit the incomplete  test suite and are not valid of inputs not included on it. 
jKali is an implementation of Kali built over Astor framework which removes code and skips the execution of code by adding \mycode{return} statements and turning True/False  expressions from \mycode{if} conditions.
As Kali, jKali does an exhaustive navigation of the search space (section \ref{sec:exhaustive}).

\subsection{jMutRepair}
\label{sec:approachjmutation}

Mutation-based repair system was introduced by Debroy et Wong \cite{debroy2010using} to repair bugs using mutation operators proposed by mutation testing approaches \cite{DeMillo1978Hint,Hamlet1977TPA}.
We implemented that system over Astor,  named jMutRepair, with has as repair operators the mutation of relational  (e.g.,$==$, $>$) and logical operators (e.g., AND, OR). 
jMutRepair does an exhaustive or selective navigation of the search space (section \ref{sec:guided_nav}).

\subsection{DeepRepair}
\label{sec:approachdeeprepair}
DeepRepair \cite{white2017dl} is an ingredient-based approach built over jGenProg that applies \emph{deep learning} based techniques during the patch synthesis process.
In particular, DeepRepair proposes new strategies for: 
\begin{inparaenum}[\it a)]
\item selecting ingredients (section \ref{sec:epingselect}) based on similarity of methods and classes: ingredients are taken from the most similar methods or classes to those that contains the bug; 
\item transforming  ingredients (section \ref{sec:epingtransf}) based on semantic of variables names: out-of-scope variables from an ingredients are replaced by in-scope variables with semantically-related names.
\end{inparaenum}

\subsection{Cardumen}
\label{sec:approachcardumen}

Cardumen \cite{CardumenArxiv} is a repair system that targets fine-grained code elements: \emph{expressions}.
It synthesizes repairs from templates mined from the applications under repair.
Then, it creates concrete patches from those templates by using a probabilistic model of variable names: it replaces template placeholders by variables with the most frequent names at the buggy location.
Cardumen explores the search space using the selective strategy (section \ref{sec:guided_nav}).

\subsection{TIBRA}
\label{sec:approachtibra}

TIBRA  ({\underline{T}}ransformed \underline{I}ngredient-\underline{B}ased \underline{R}epair \underline{A}pproach), is an extension of jGenProg which, as difference with the original GenProg who discards ingredients with out-of-scope variables, it applies transformation into the ingredients.
The approach  adapts an ingredient i.e., statement taken somewhere, by replacing all variables out-of-scope from it by in-scope  variables randomly chosen from the buggy location.

\section{Evaluation}
\label{sec:evaluation}

In this section we present an evaluation of repair approaches built over Astor.
We first study the capacity of approaches presented in section \ref{sec:approachessummary} to repair real bugs. 
Then, we focus on the \emph{ingredient pool}, the component of GenProg that stores the source code used for synthesizing candidate patches.
There, we compare different implementations of the extension points from section \ref{sec:extensionpoints} related to the creation of the ingredient pool and to the transformation of ingredients.

\subsection{Research questions}
The research questions that guide the evaluation of Astor are:

\newcommand{\rqdj}{RQ 1.1 - How many bugs from Defects4J are repaired using the repair approaches built over Astor?}

\newcommand{\rquniques}{RQ 1.2: What are the bugs uniquely repaired by approaches from Astor?}

\newcommand{\rqtarget}{RQ 1.3 - Which  code granularity repairs more bugs?}

\newcommand{\rqisp}{
RQ 2.1 -To what extent does a reduced ingredient space impact repairability?
}

\newcommand{\rqit}{RQ 2.2 - To what extent does the ingredient transformation strategy impact repairability?}

\begin{enumerate}[\it 1)]

\item Repairability:

\item[] 
\begin{enumerate}
\item[]{\rqdj}
\item[]{\rquniques}

\item[]{\rqtarget}

\end{enumerate}

\item Focus on ingredient-based repair:
\item[] 
\begin{enumerate}
\item[]{\rqisp}

\item[]{\rqit}
\end{enumerate}
\end{enumerate}

\subsection{Protocol}

We have used repair approaches from Astor for a large evaluation consisting in searching for patches for 357 bugs from the Defects4J benchmarks \cite{JustJE2014}, those from 5 projects: Apache Commons Math, Apache Commons Lang, JFreeChart, Closure and Joda Time.
A patch is said to be \emph{test-adequate} if it passes all tests, including the failing one. 
As shown by previous work \cite{Qi2015}, a patch may be test-adequate yet incorrect, when it only works on the inputs from the test suite and does not generalize. 
Those patches are known as \emph{overfitting} patches \cite{smith2015cure,Le:2018:OSA,Yu2018}.
This paper does not aim at studying the correctness of the test-suite adequate patches found by a repair approach as done in, for instance, in \cite{defects4j-repair}. 
In the rest of the paper, \emph{``to repair a bug''} means to find a test-suite adequate patch.

The main experimental procedure is composed of the following steps.
First, we selected the repair approaches to study according to the addressed research question.
Then, we  created scripts for launching repairs attempts for each approach on each bug from Defects4J.
A repair attempt consists on the execution of a repair approach executed with a timeout of 3 hours and a concrete value of random seed.
In case of exhaustive approach such as jKali, the seed is not used due it does not have any stochastic sub-component.
For each configuration, we run at least 3 repair attempts (i.e., 3 different seeds).
Finally, we collected the results from each repair attempt, grouped the results (i.e., patches and statistics) according to the configuration, and compared the results.

\subsection{Evaluation of Repairability}

In this section we focus on the ability of repair approaches  over Astor introduced in section \ref{sec:approachessummary} to repair bugs from Defects4J and we compare their repairability against other repairs systems.

\subsubsection{\rqdj} 

Table \ref{tab:d4j_all_repaired} displays the bugs from Defects4J repaired by approaches built over Astor framework. 
In total, 98 bugs out of 357 (27.4\%) are repaired by at least one repair approach.
Six approaches were executed: jGenProg (49 bugs repaired),
jKali (29), jMutRepair (23),  DeepRepair (51), Cardumen (77),  and TIBRA (35).

We observe that there are 9 bugs such as Chart-1 and Math-2 that all evaluated  repair approaches found at least one test-suite adequate patches.
Contrary, 35 bugs were repaired by only one approach.  
19 of them, such as Math-101, are repaired by Cardumen, 9 only by DeepRepair (e.g., Math-22), 3 by jGenProg (e.g., Chart-19), 3 by TIBRA (e.g., Chart-23), and one for jMutRepair (Closure-38).

\newcommand{\repairedsymbol}{R}
\begin{table}[!t]
\renewcommand{\arraystretch}{1.1}
\tiny
\caption{Bugs from dataset Defects4J repaired by approaches built over Astor.
In total, 98 bugs from 5 Java projects were repaired. \repairedsymbol{} means `bug with at least one test-suite adequate patch'.
}
\begin{tabular}{ll}
\begin{tabular}{|l|c||l|l|l|l|l|l|c|}
\hline

\rot{Project}  &  \rot{Bug Id}  &  \rot{jGenProg} &  \rot{jKali}  &  \rot{jMutRepair}  &  \rot{DeepRepair}  &  \rot{Cardumen}  &  \rot{TIBRA}  & { \rot{\#  approaches}} \\ 
\hline 
\hline
\multirow{19}{*}{\rotatebox[origin=c]{90}{Chart}} 
  & 1  &  \repairedsymbol  & \repairedsymbol & \repairedsymbol & \repairedsymbol & \repairedsymbol & \repairedsymbol & 6 \\ 
  &  3  & \repairedsymbol &    &    & \repairedsymbol & \repairedsymbol & \repairedsymbol & 4 \\ 
  &  4  &    &    &    &    & \repairedsymbol &    & 1 \\ 
  &  5  & \repairedsymbol & \repairedsymbol &    & \repairedsymbol & \repairedsymbol & \repairedsymbol & 5 \\ 
  &  6  &    &    &    &    & \repairedsymbol & \repairedsymbol & 2 \\ 
  &  7  & \repairedsymbol &    & \repairedsymbol & \repairedsymbol & \repairedsymbol & \repairedsymbol & 5 \\ 
  &  9  &    &    &    & \repairedsymbol & \repairedsymbol &    & 2 \\ 
  &  11  &    &    &    &    & \repairedsymbol & \repairedsymbol & 2 \\ 
  &  12  & \repairedsymbol &    &    & \repairedsymbol & \repairedsymbol & \repairedsymbol & 4 \\ 
  &  13  & \repairedsymbol & \repairedsymbol &    & \repairedsymbol & \repairedsymbol &    & 4 \\ 
  &  14  &    &    &    & \repairedsymbol &    &    & 1 \\ 
  &  15  & \repairedsymbol & \repairedsymbol &    & \repairedsymbol & \repairedsymbol &    & 4 \\ 
  &  17  &    &    &    &    & \repairedsymbol & \repairedsymbol & 2 \\ 
  &  18  &    &    &    & \repairedsymbol &    &    & 1 \\ 
  &  19  & \repairedsymbol &    &    &    &    &    & 1 \\ 
  &  23  &    &    &    &    &    & \repairedsymbol & 1 \\ 
  &  24  &    &    &    &    & \repairedsymbol & \repairedsymbol & 2 \\ 
  &  25  & \repairedsymbol & \repairedsymbol & \repairedsymbol & \repairedsymbol & \repairedsymbol &    & 5 \\ 
  &  26  & \repairedsymbol & \repairedsymbol & \repairedsymbol & \repairedsymbol & \repairedsymbol & \repairedsymbol & 6 \\ 
\hline 
\multirow{13}{*}{\rotatebox[origin=c]{90}{Closure}}
  &  7  &    &  \repairedsymbol  &    &    & \repairedsymbol &    & 2 \\ 
  &  10  &    &   \repairedsymbol & \repairedsymbol   &    & \repairedsymbol &    & 2 \\ 
  &  12  &    &    &    &    & \repairedsymbol &    & 1 \\ 
  &  13  &    &   \repairedsymbol &    &    & \repairedsymbol &    & 2 \\ 
  &  21  & \repairedsymbol &  \repairedsymbol  & \repairedsymbol   &    & \repairedsymbol &    & 3 \\ 
  &  22  & \repairedsymbol &  \repairedsymbol  & \repairedsymbol  &    & \repairedsymbol &    & 4 \\ 
  &  33  &    &    &    &    & \repairedsymbol &    & 1 \\ 
  &  38  &    &    & \repairedsymbol   &    &   &    & 1 \\ 
  &  40  &    &    &    &    & \repairedsymbol &    & 1 \\ 
  &  45  &    &  \repairedsymbol  &    &    & \repairedsymbol &    & 2 \\ 
  &  46  & \repairedsymbol &   \repairedsymbol &    &    & \repairedsymbol &    & 3 \\ 
  &  49  & \repairedsymbol &    &    &    &    &    & 1 \\ 
  &  55  &    &    &    &    & \repairedsymbol &    & 1 \\ 
  &  133  &    &    &    &    & \repairedsymbol &    & 1 \\ 
\hline 
\multirow{9}{*}{\rotatebox[origin=c]{90}{Lang}} 
  &  7  & \repairedsymbol &    &    & \repairedsymbol & \repairedsymbol &    & 3 \\ 
  &  10  & \repairedsymbol &    &    & \repairedsymbol & \repairedsymbol &    & 3 \\ 
  &  14  &    &    &    &    & \repairedsymbol &    & 1 \\ 
  &  20  &    &    &    & \repairedsymbol &    &    & 1 \\ 
  &  22  & \repairedsymbol &    &    & \repairedsymbol & \repairedsymbol &    & 3 \\ 
  &  24  & \repairedsymbol &    &    & \repairedsymbol & \repairedsymbol &    & 3 \\ 
  &  27  & \repairedsymbol &    & \repairedsymbol & \repairedsymbol & \repairedsymbol &    & 4 \\ 
  &  38  &    &    &    & \repairedsymbol &    &    & 1 \\ 
  &  39  & \repairedsymbol &    &    & \repairedsymbol & \repairedsymbol &    & 3 \\ 
\hline 
\multirow{8}{*}{\rotatebox[origin=c]{90}{Math}} 
  &  2  & \repairedsymbol & \repairedsymbol & \repairedsymbol & \repairedsymbol & \repairedsymbol & \repairedsymbol & 6 \\ 
  &  5  & \repairedsymbol &    &    & \repairedsymbol & \repairedsymbol & \repairedsymbol & 4 \\ 
  &  6  &    &    &    & \repairedsymbol & \repairedsymbol &    & 2 \\ 
  &  7  & \repairedsymbol &    &    &  &    &    & 1 \\ 
  &  8  & \repairedsymbol & \repairedsymbol &    & \repairedsymbol & \repairedsymbol & \repairedsymbol & 5 \\ 
  &  18  &    &    &    & \repairedsymbol & \repairedsymbol &    & 2 \\ 
  &  20  & \repairedsymbol & \repairedsymbol   &    & \repairedsymbol & \repairedsymbol & \repairedsymbol & 5 \\ 
  &  22  &    &    &    & \repairedsymbol &    &    & 1 \\ 
  \hline
\end{tabular}

\begin{tabular}{|l|c||l|l|l|l|l|l|c|}
\hline
\rot{Project}  &  \rot{Bug Id}  &  \rot{jGenProg} &  \rot{jKali}  &  \rot{jMutRepair}  &  \rot{DeepRepair}  &  \rot{Cardumen}  &  \rot{TIBRA}  & \multicolumn{1}{l}{\rot{\# approaches} } \\ 
\hline 
\hline 
\multirow{41}{*}{\rotatebox[origin=c]{90}{Math}} 
  &  24  &    &    &    & \repairedsymbol &    &    & 1 \\ 
  &  28  & \repairedsymbol & \repairedsymbol & \repairedsymbol & \repairedsymbol & \repairedsymbol & \repairedsymbol & 6 \\ 
  &  30  &    &    &    &    & \repairedsymbol &    & 1 \\ 
  &  32  & \repairedsymbol & \repairedsymbol &    & \repairedsymbol & \repairedsymbol & \repairedsymbol & 5 \\ 
  &  33  &    &    &    &    & \repairedsymbol &    & 1 \\ 
  &  39  & \repairedsymbol &    &    &    &    &    & 1 \\ 
  &  40  & \repairedsymbol & \repairedsymbol & \repairedsymbol & \repairedsymbol & \repairedsymbol &    & 5 \\ 
  &  41  &    &    &    &    & \repairedsymbol &    & 1 \\ 
  &  44  & \repairedsymbol   &    &    &  &    & \repairedsymbol & 2 \\ 
  &  46  &    &    &    &    & \repairedsymbol &    & 1 \\ 
  &  49  & \repairedsymbol & \repairedsymbol &    & \repairedsymbol & \repairedsymbol & \repairedsymbol & 5 \\ 
  &  50  & \repairedsymbol & \repairedsymbol & \repairedsymbol & \repairedsymbol & \repairedsymbol & \repairedsymbol & 6 \\ 
  &  53  & \repairedsymbol &    &    & \repairedsymbol &    & \repairedsymbol & 3 \\ 
  &  56  & \repairedsymbol &    &    & \repairedsymbol &    &    & 2 \\ 
  &  57  &    &    & \repairedsymbol & \repairedsymbol & \repairedsymbol &    & 3 \\ 
  &  58  &    &    & \repairedsymbol & \repairedsymbol & \repairedsymbol &    & 3 \\ 
  &  60  & \repairedsymbol &    &    & \repairedsymbol & \repairedsymbol &    & 3 \\ 
  &  62  &    &    &    &    & \repairedsymbol &    & 1 \\ 
  &  63  &    &    &    & \repairedsymbol & \repairedsymbol & \repairedsymbol & 3 \\ 
  &  64  &  \repairedsymbol  &    &    &  &    &    & 1 \\ 
  &  69  &    &    &    &    & \repairedsymbol &    & 1 \\ 
  &  70  & \repairedsymbol &    &    & \repairedsymbol & \repairedsymbol &    & 3 \\ 
  &  71  & \repairedsymbol &    &    & \repairedsymbol &    &    & 2 \\ 
  &  72  &    &    &    &    &    & \repairedsymbol & 1 \\ 
  &  73  & \repairedsymbol &    &    & \repairedsymbol & \repairedsymbol & \repairedsymbol & 4 \\ 
  &  74  & \repairedsymbol &    &    & \repairedsymbol & \repairedsymbol &    & 3 \\ 
  &  77  &    &    &    & \repairedsymbol &    &    & 1 \\ 
  &  78  & \repairedsymbol & \repairedsymbol &    & \repairedsymbol & \repairedsymbol & \repairedsymbol & 5 \\ 
  &  79  &    &    &    &    & \repairedsymbol & \repairedsymbol & 2 \\ 
  &  80  & \repairedsymbol & \repairedsymbol &  \repairedsymbol  & \repairedsymbol & \repairedsymbol & \repairedsymbol & 6 \\ 
  &  81  & \repairedsymbol & \repairedsymbol & \repairedsymbol & \repairedsymbol & \repairedsymbol & \repairedsymbol & 6 \\ 
  &  82  & \repairedsymbol & \repairedsymbol & \repairedsymbol & \repairedsymbol & \repairedsymbol & \repairedsymbol & 6 \\ 
  &  84  & \repairedsymbol & \repairedsymbol & \repairedsymbol & \repairedsymbol & \repairedsymbol &    & 5 \\ 
  &  85  & \repairedsymbol & \repairedsymbol &   \repairedsymbol & \repairedsymbol & \repairedsymbol & \repairedsymbol & 6 \\ 
  &  88  &    &    & \repairedsymbol &    & \repairedsymbol &    & 2 \\ 
  &  95  & \repairedsymbol & \repairedsymbol &    &    & \repairedsymbol & \repairedsymbol & 4 \\ 
  &  97  &    &    &    &    & \repairedsymbol & \repairedsymbol & 2 \\ 
  &  98  &    &    &    & \repairedsymbol &    &    & 1 \\ 
  &  101  &    &    &    &    & \repairedsymbol &    & 1 \\ 
  &  104  &    &    & \repairedsymbol   &    & \repairedsymbol &    & 2 \\ 
  &  105  &    &    &    &    & \repairedsymbol &    & 1 \\ 
\hline 
\multirow{7}{*}{\rotatebox[origin=c]{90}{Time}}
  &  4  & \repairedsymbol &    &    &    & \repairedsymbol & \repairedsymbol & 3 \\ 
  &  7  &    &    &    &    & \repairedsymbol &    & 1 \\ 
  &  9  &    &    &    &    & \repairedsymbol &    & 1 \\ 
  &  11  & \repairedsymbol & \repairedsymbol   & \repairedsymbol &    & \repairedsymbol & \repairedsymbol & 5 \\ 
  &  17  &    &    &    &    & \repairedsymbol &    & 1 \\ 
  &  18  &    &    &    &    & \repairedsymbol &    & 1 \\ 
  &  20  &    &    &    &    &    & \repairedsymbol & 1 \\ 
\hline 
\hline 
 {\tiny{$\sum$}}  &  {98}  &  {49}  &  {29}  &  {23}  &  {51}  &  {77}  &  {35}  & \\ 
 \hline
\end{tabular}

\end{tabular}
\label{tab:d4j_all_repaired}
\end{table}

\begin{framed}
{\bf Response to RQ 1.1:}
The repair approaches built over Astor find test-adequate patches for {\bf{98}} real bugs out of {357} bugs from Defects4J.
The best approach is Cardumen: it finds a test-suite adequate patch for 77 bugs.
\end{framed}

Compared with other repair system evaluated over Defects4J, approaches from Astor repair more bugs (98 bugs repaired) from Defects4J than:  ssFix \cite{xin2017leveraging} (60 bugs repaired), ARJA \cite{Arja1712.07804} (59 bugs),   ELIXIR \cite{Saha:2017:EEO} (40 bugs),  GP-FS \cite{gpfl2017} (37 bugs), JAID \cite{Chen2017CPR} (31 bugs), ACS \cite{Xiong2017} (18 bugs),  HDRepair \cite{le2016history} (15 bugs from \cite{xin2017leveraging}). 
In particular, Cardumen repairs more that all those approaches: it found test-suite adequate patches for 77 bugs.
On the contrary, Nopol \cite{nopol} (103 bugs repaired \cite{durieux:hal-01480084}) repairs more than Astor framework.

\subsubsection{\rquniques}

We consider automated program repair approaches from the literature for which:
\begin{inparaenum}[\it 1)]
\item the evaluation was done over the dataset Defects4J; 
\item the identifiers of the repaired bugs from Defect4J are given on the respective paper or included in the appendix.
\end{inparaenum}

We found 10 repair systems that fulfill both criteria:  Nopol \cite{nopol,durieux:hal-01480084}, jGenProg \cite{defects4j-repair}, DynaMoth \cite{Durieux2016DDC}, HDRepair \cite{le2016history}, DeepRepair \cite{white2017dl}, ACS \cite{Xiong2017}, GP-FS \cite{gpfl2017}, 
JAID \cite{Chen2017CPR},  ssFix \cite{xin2017leveraging} and ARJA \cite{Arja1712.07804}.
In the case of HDRepair, as  neither the  identifiers of the repaired bugs nor the actual patches were reported by \cite{le2016history}, we considered the results reported by \cite{xin2017leveraging} (ssFix's authors) who executed the approach.
We discarded the Java systems  JFix (\cite{Le2017JSR}), S3 (\cite{Le2017SSS}), Genesis (\cite{Long2017AIC}) (evaluated over different bug datasets) and ELIXIR \cite{Saha:2017:EEO} (repaired ids from Defect4J not publicly available).

Approaches built on Astor framework found test-suite adequate patches for {\bf \uniquerepaired{}} new bugs  of Defects4J, for which no other system ever has managed to find a single one.
Those  uniquely repaired bugs are: 
5 from Math (ids: 62, 64, 72, 77, 101),
2 from Time (9, 20),
2 from Chart (11, 23), and 
2 from Closure (13, 46).

\begin{framed}
{\bf Response to RQ 1.2:}
The repair approaches built over Astor find new unique test-adequate patches. Astor repair  {\bf{\uniquerepaired{}}} bugs which have never been repaired previously by any repair system.
\end{framed}

In the remain of the evaluation section, we evaluates different section implementation for tree extension points from \ref{sec:extensionpoints}.

\subsubsection{\rqtarget}

We compared the repairability of two approaches that use different implementations of the extension point \epmpgranularity{} for manipulating different granularity of code source.
The level of code granularity impacts on the size of the search space and thus in the ease to find a patch.
On one hand, Cardumen approach is able to synthesize a fine-grained patches by modifying, for example, an expression inside a statement with other expressions inside.
On the other hand, the code modifications done by jGenProg (or by any approach that extends it such as DeepRepair or TIBRA) to find a patch are coarse-level: as it works at the level of \emph{statements}, an entire statement is inserted, deleted or replaced.

Table \ref{tab:d4j_all_repaired} shows that 
Cardumen (expression level) and jGenProg's family approaches (statement level)
repaired 77 and 72 bugs, respectively, and 52 of them were repaired by both approaches.
Even the difference of repairability is not statistically significant enough (Mann-Whitney  test shows a p-value of 0.649), 
the experiment shows that there is a considerable portion of bugs that can be repaired only in a given granularity: 25 bugs are repaired by Cardumen but not by jGenProgs family, and 20 bugs are repaired by only this latter family of approaches.

\begin{framed}
{\bf Response to RQ 1.3:}
The extension point \epmpgranularity{} has an impact on repair.
We compared the extensions \emph{statements} and \emph{expression} implemented in jGenProg and Cardumen, respectively.
By applying operators at the level of statements and expressions, \bold{52} patches (72.2\% and 67.5\%, resp.) are repaired by both jGenProg and Cardumen, respectively.
The remaining bugs (20 and 25, resp.) are only repaired using a specific granularity.

\end{framed}

The implication is that, for those bugs repaired by both approaches, there are patches that:
\begin{inparaenum}[\it a)]
\item produce similar behaviours w.r.t the test-suite (i.e., passing all tests), and 
\item the changed codes have different granularities.  
\end{inparaenum}
For example, both jGenProg and Cardumen synthesize the following patch for bug Math-70:

\begin{lstlisting}[caption={Patch for Math-70 at class BisectionSolver.java},basicstyle=\small]
72 -		return solve(min, max);
72 +		return solve(f, min, max);
 	}
\end{lstlisting}

jGenProg synthesizes that patch by applying the replace operator to the modification point that references to the \mycode{return} \emph{statement} at line 72 of class BisectionSolver. The replacement \mycode{return} statement (i.e., the ingredient) is taken from line 59 from the same buggy class.
Meanwhile, Cardumen arrives to the same patch by replacing a modification point that references to the \emph{expression} corresponding to the method invocation \mycode{solve} inside the \mycode{return} statement at line 72. The replacement synthesized from a template: 

\mycode{solve(_UnivariateRealFunction_0, _double_1, _double_2)} mined from the same class  BisectionSolver and instantiated using variables in scope at line 72.

One of the 25 bugs repaired by Cardumen but not by jGenProg is Math-101
The patch proposed modifies the expression related to a variable initialization (endIndex) at line 376 from \mycode{startIndex + n} to \mycode{source.length()} on class ComplexFormat.

\begin{lstlisting}[caption={Patch for Math-101 by Cardumen},basicstyle=\small]
376 -	int endIndex = startIndex + n;
376 +	int endIndex = source.length();
\end{lstlisting}

Cardumen is able to synthesize the patch by instantiating the template \mycode{\_String\_0.length()} mined from the application under repair.
Approaches working at a different (and coarse) granularity are not capable to synthesize that patch: in the case of jGenProg, the statement \mycode{int endIndex = source.length();}  (the ingredient for the fix) does not exist anywhere in the application under repair.

Astor framework provides  to developers of approaches the flexibility to manipulate specific code elements at a given granularity level by  implementing the extension point \epmpgranularity.
For instance, jMutRepair implements \epmpgranularity{} to manipulate relational and logical binary operator. This allows to target to specific defect classes and to reduce the search space size.
That is the case for bug Closure-38, which is only repaired by jMutRepair.

\begin{lstlisting}[caption={Patch for Closure-38 by jMutRepair at class CodeConsumer},basicstyle=\small]
245 - if (x < 0) && (prev == '-') {
245 + if (x <= 0) && (prev == '-') {
\end{lstlisting}

\subsection{Design of ingredient-based repair approaches}

In this section we study and compare two different implementations for the extension points related to ingredient-based repair approaches \epingredientpool{} and \epingredienttransformation{}.
The goal of the next experiments is to study whether improvements introduced via those extension points impact on the repairability with respect to the vanilla jGenProg.

\subsubsection{\rqisp}

We evaluate the extension point \epingredientpool{} by using different strategies for building an ingredient pool (Section \ref{sec:epingspacedef}).
The experiment's goal is to know whether using a reduced ingredient pool, such as File and Package pools preserves the repair capability of a the vanilla GenProg approach (which uses Global scope as default).

To study the impact of using a reduced ingredient pool, 
we executed jGenProg on Defects4J using the baseline ingredient pool used by the original GenProg \cite{Weimer2009} (i.e, Global pool) and the optimized modes (File and Package pool), based on the empirical evidences of ingredient's locations \cite{martinez2014icse,Barr2014PSH}.
For each bug from Defects4J and for each pool type (File, Package and Global) we executed 3 repairs attempts with a timeout of 3 hours, using on each attempt a different random seed value.

In this experiment, jGenProg repaired, in total, 39 bugs using ingredient based repair operators.\footnote{In this experiment we did not consider all bugs repaired using the remove operator.} 
The numbers of repaired bugs are different  according with the ingredient pool used by jGenProg:
it repaired 33, 28 and 14 bugs using File, Package and Global pool, respectively.
We applied the Wilcoxon rank sum test (aka Mann-Whitney test) to verify that the difference of repairability between  jGenProg using a reduced space and jGenProg using Global are statically significant, obtaining p-values of 1.296e-05 (File vs Global) and 0.001613 (Package vs Global).

Moreover, we found that \emph{all} the bugs repaired using Global scope were also repaired by jGenProg  using either File or Package scopes.
This means that a reduced ingredient space still continue having, at least, one ingredient that is used to synthesize a test-suite adequate patch.

The reduction of the ingredient space produces another advantage: it allows jGenProg to find faster the first test-suite adequate patch.
For the 12 bugs that were repaired jGenProg using both File and Global scopes, 10 of them were repaired faster using File scope (83\%), saving on average 22.8 minutes.
Similarly, for the 14 bugs that were repaired using both Package and Global scopes, 12 of them were repaired faster using package.
This result validates the fact that locality-aware repair speeds-up repair \cite{Barr2014PSH,martinez2014icse}.

\begin{framed}
{\bf Response to RQ 2.1:} The extension point \epingredientpool{} impacts the numbers of repaired bugs and on the repair time.
We compared jGenProg  using the File, Package and Global (default by GenProg) scopes for building the ingredient pool.
For our repair attempts bounded by a 3-hours maximum budget, the File and Package ingredients pools found more test-suite adequate patches and faster (reduction of 22 minutes on average) than the baseline (Global).

\end{framed}

\subsubsection{\rqit}
\label{sec:deep2}

By default, the original GenProg does not apply any transformation of ingredients once they are selected from the ingredient search space.
As we presented in Section \ref{sec:epingtransf} Astor provides an extension point \epingredienttransformation{} for plugging an ingredient transformation strategy.
 We now present the evaluation of two different implementations of \epingredienttransformation{}, both included in Astor.

First, we evaluated the approach TIBRA (Section \ref{sec:approachtibra}), which replaces out-of-scope variables from an ingredient, and we compared the results against those from the original jGenProg (which does not transformation ingredients selected from the search space).
TIBRA repaired 11 bugs (e.g., Math-63 in Table \ref{tab:d4j_all_repaired}) that jGenProg did not repair, showing that the transformation of ingredient allows to find patches for unrepaired bugs.
The Wilcoxon signed-rank test shows that the difference between the repairability from vanilla jGenProg and TIBRA is statistically significant, with a p-value of 1.247e-09.

A second implementation of the extension point  \epingredienttransformation{}  is a cluster-based ingredient transformation strategy
proposed by DeepRepair \cite{white2017dl} (which is built over Astor).
We carried out a second experiment that compared jGenProg using this strategy (named DeepRepair RE in \cite{white2017dl}) with the original jGenProg.
The results showed that there is not a statistical differences between the repairability of bugs from Defects4j.\footnote{Wilcoxon test shows a p-value = 0.3626. Consequently, in \cite{white2017dl} we fail to reject the Null Hypothesis that states that the DeepRepair's ingredient transformation strategy RE generates the same number of test-adequate patches as jGenProg.}
The transformation  of ingredients using the cluster-based strategy allows to repair only 4  bugs that jGenProg could not (e.g., Math-98 in Table \ref{tab:d4j_all_repaired}).
However, we found that there are notable differences between DeepRepair and jGenProg patches: 53\%, 3\%, and 53\% of DeepRepair’s patches for Chart, Lang, and Math, respectively, are not found by jGenProg.

Finally, we compared TIBRA and DeepRepair.
TIBRA repairs  in total 35 bugs, of them  21 are also repaired by DeepRepair.
This means that 28 and 14 bugs are only repaired by DeepRepair (RE) and TIBRA, respectively. 
The Wilcoxon signed-rank test shows that the difference between the repairability from DeepRepair (RE) and TIBRA is statistically significant, with a p-value of 0.02739.
This last experiment shows the benefits of using a customized strategy based on cluster of variables names over a strategy based on random selection of variables.

\begin{framed}
{\bf Response to RQ 2.2:}
The extension point \epingredienttransformation{} impacts the repairability.
We compared three extensions: \emph{no-transformation}, \emph{cluster-based}, and \emph{random-variable-replacement} implemented in  jGenProg, DeerRepair and TIBRA, respectively.
DeepRepair and TIBRA  discover new test-suite adequate patches that cannot by synthesized by jGenProg for 4 and 11 bugs, respectively.
\end{framed}

\section{Related Work}
\label{sec:relatedwork}

\subsection{Program Repair Frameworks}

To our knowledge, Astor is the first and unique framework on Java that implements a repair workflow from generate-and-validate repair approaches and provides twelve extension points for which the program repair researcher can either choose an existing component (i.e., taking one already implemented design decision in the design space), or can implement a new technique.

\subsection{Works that extend approaches from Astor}

In this section we present the repair approaches and extensions from the bibliography that were built over the Astor framework.
Tanikado et al.  \cite{Tanikado2017NewStrategies} extended jGenProg provided by Astor framework for introducing two novel strategies. One, named   \emph{similarity-order}, which extends extension point \epingredientselection, chooses ingredients according to code fragment similarities. The second one, named \emph{freshness-order}, which extends the  modification point \epmpselection, consists on selecting, with a certain priority, modification points whose statements were more recently updated.
Wen et al. \cite{gpfl2017}  presented a systematic empirical study that explores the influence of fault space on search-based repair techniques.  For that experiment, the author created the approach GP-FS, an extension of jGenProg, which receives as input a faulty space. In their experiment, the authors generated several fault spaces with different accuracy,  finding that GP-FS is capable of fixing more bugs correctly when fault spaces with high accuracy are fed.
White et al. \cite{white2017dl} presented DeepRepair, an extension of jGenProg, which navigates the search space guided by method and class similarity measures inferred with deep unsupervised learning.  DeepRepair was incorporated to Astor framework as built-in approach. 

\subsection{Works that execute built-in approaches from  Astor}

Works from the literature executed repair approaches from Astor framework during the evaluation of their approaches.
For example, 
Yuefei  presents and study \cite{LiuYuefei2017} for  understanding and generating patches for bugs introduced by
third-party library upgrades.
The author run jGenProg from Astor to repair the 6 bugs, finding correctly 2 patches for bugs, and a  test-suite adequate but yet incorrect patch for another bug.
The approach ssFix \cite{xin2017leveraging}  performs syntactic code search to find existing code from a code database (composed by the application under repair and external applications) that is syntax-related to the context of a bug statement.
In their evaluation, the authors executed two approaches from Astor,  jGenProg  and  jKali, using the same machines and configuration that used for executing ssFix.

\subsection{Works that compare repairability against that one from  built-in approaches from  Astor}

We have previously   executed jGenProg and jKali over bugs from Defects4J \cite{JustJE2014} and analyzed the correctness of the generated patches \cite{defects4j-repair}. 
Note that, the number of repaired bugs we reported in that experiment, executed  in 2016, are lower that the results we present in this paper in section \ref{sec:evaluation}. The main reason is we have applied several improvements and bugfixings over Astor framework since that experiment.

Other works have used the mentioned evaluation of jGenProg and jKali presented in \cite{defects4j-repair}  for measuring the improvement introduced by their new repair approaches.
For example, Le et al. presented  a new repair approach named HDRepair \cite{le2016history} which  leverages on the development history to effectively guide and drive a program repair process. The approach first mines bug fix patterns from the history of many projects and the then employ existing mutation operators to generate fix candidates for a given buggy program.
The approach ACS (Automated Condition Synthesis) \cite{Xiong2017}, targets to insert or modify an “if” condition to repair defects by combining three heuristic ranking techniques that exploit 1) the structure of the buggy program, 2) the document of the buggy program (i.e., Javadoc comments embedded in the source code), and 3) the conditional expressions in existing projects.
Yuan and Banzhaf \cite{Arja1712.07804} present ARJA, a genetic-programming based repair approach for automated repair of Java programs. ARJA introduces a test filtering procedure that can speed up the fitness evaluation and three types of rules that can be applied to avoid unnecessary manipulations of the code. ARJA also considers the different representation of ingredient pool introduced by Astor framework \cite{astor2016}.
In addition to the evaluation of Defects4J, the authors evaluated the capacity of repair real multi-location bugs over another dataset built by themselves.
Saha et al. presented Elixir \cite{Saha:2017:EEO}  a  repair technique which has a 
fixed set of parameterized program transformation schemas used for synthesized candidate patches. 
JAID by \cite{Chen2017CPR} is a state-based dynamic program analyses which synthesizes patches based on schemas (5 in total).
Each schema trigger a fix action when a suspicious state in the system is reached during a computation. JAID has 4 types of fix actions, such as modify the state directly by assignment, and affect the state that is used in an expression.

\subsection{Works that analyze patches from  built-in approaches from  Astor}

Other works have analyzed the publicly available patches of jGenProg and jKali from our previous evaluation of repair approaches over Defects4J dataset \cite{defects4j-repair}.
Motwani et al. \cite{motwani2017automated} analyzed the characteristics of the defects that repair approaches (including jGenProg and jKali) can repair.
They found that automated repair techniques are less likely to produce patches for defects that required developers to write a lot of code or edit many files. 
They found that the  approaches that target Java code, such as those from Astor, are more likely to produce patches for high-priority defects than the techniques which target C code. 
Yokoyama et al. \cite{Yokoyama2017Evaluating} extracted characteristics of defects from defect reports such as priority and evaluated the performance of repairs approaches  against 138 defects in open source Java project included in Defects4J.
They found that jGenProg is able to find patch for  many high-priority defects (1 Blocker, 2 Critical, and 11 Major).
Liu et al \cite{liu2017identifying} presented a approach that heuristically determines the correctness of the generated patches, by exploiting the behavior similarity of test case executions.
The approach  is capable of automatically detecting  as incorrect the 47.1\% and 52.9\% of patches from  jGenprog and jKali, respectively.
Jiang et al. \cite{jiang2017can} analyzed the Defects4J dataset for finding bugs with weak test cases.
They results shows that 42 (84.0\%) of the 50 defects could be fixed with weak test suites, indicating that, beyond the current techniques have a lot of rooms for improvement, weak test suites may not be the key limiting factor for current techniques.

\subsection{Other test-suite based repair approaches}

During the last decade, other approaches  target other programming languages (such as C) or we evaluated over other datasets rather than Defects4J were presented.
Arcuri \cite{ArcuriEvolutionary} applies co-evolutionary computation to automatically generate bug fixes for Java program.
GenProg \cite{Weimer2009,LeGoues2012TSEGP}, one of the earliest generate-and-validate techniques, uses genetic programming to search the repair space and generates patches created from existing code from elsewhere in the same program. It has three repair operators: add, replace or remove statements.
Other approaches have extended GenProg: for example, AE  \cite{weimer2013AE} employs a novel deterministic search strategy and uses program equivalence relation to reduce the patch search space.
The original implementation  \cite{Weimer2009} targets C code and was evaluated against dataset with C bugs such as ManyBugs and IntroClass \cite{LeGoues2015MB}.
Astor provides a Java version of GenProg called jGenProg which also employs genetic programming for navigating the search space.
RSRepair  \cite{rsrepair} has the same search space as GenProg but uses random search instead, and the empirical evaluation shows that random search can be as effective as genetic programming.
Astor is able to execute a Java version of RSRepair by choosing random strategies for the selection of modification points (extension point \epmpselection) and operators (extension point \epmoperatorselection).
Debroy \& Wong \cite{debroy2010using} propose a mutation-based repair method inspired from mutation testing. 
This work combines fault localization with program mutation to exhaustively explore a space of possible patches.
Astor includes a Java version of this approach called jMutRepair.
Kali  \cite{Qi2015} has recently been proposed to examine the fixability power of simple actions, such as statement removal. As GenProg, Kali targets C code. Astor proposes a Java version of Kali,  which includes all transformations proposed by Kali.

Other approaches have proposed new set of  repair operators. For instance, 
PAR \cite{Kim2013}, which shares the same search strategy with GenProg, uses  patch templates derived from human-written patches to construct the search space.
The PAR tool used  the original evaluation is not publicly available.  However, it is possible to implement PAR over the Astor framework by implementing the repair operator based on those templates using the extension point \epmoperatordef .
The approach SPR  \cite{spr} uses a set of predefined transformation schemas to construct the search space, and patches are generated by instantiating the schemas with condition synthesis techniques.  SPR  is publicly available but targets C programs.
An extension of SPR, Prophet \cite{prophet} applies probabilistic models of correct code learned from successful human patches to prioritize candidate patches so that the correct patches could have higher rankings.

There are approaches that leverage on human written bug fixes. 
For example, 
Genesis \cite{Long2017AIC} automatically infers code transforms for automatic patch generation. The code transformation used Genesis are automatically infer from previous successful patches.
The approach first mines bug fix patterns from the history of many projects and the then employ existing mutation operators to generate fix candidates for a given buggy program.
Both approaches need as input, in addition to the buggy program and its test suite, a set of bug fixes.
Two approaches leveraged on semantics-based examples.
SearchRepair  \cite{Ke2015RPS} uses a large database of human-written code fragments encore as satisfiability modulo theories (SMT) constraints on their input-output behavior for synthesizing candidates repairs.
S3 (Syntax- and Semantic-Guided Repair Synthesis)  by \cite{Le2017SSS}, a repair synthesis engine that leverages programming-by-examples methodology to synthesize repairs.

Other approaches belong to the family of \emph{synthesis-based repair approaches}. For example, SemFix \cite{Nguyen:2013:SPR} is a constraint based repair approach for C. This approach provides patches for assignments and conditions by combining symbolic execution and code synthesis. 
Nopol  \cite{nopol} is also a constraint based method, which focuses on fixing bugs in  if  conditions and missing preconditions, as Astor, it is implemented for Java and publicly available. 
DynaMoth \cite{Durieux2016DDC} is based on Nopol, but replaces the   SMT-based synthesis component of Nopol by a new synthesizer, based on dynamic exploration,  that is able to generate richer patches than Nopol e.g., patches on \emph{If} conditions  with method invocations inside their condition.
DirectFix   \cite{directfix} achieves the simplicity of patch generation with a Maximum Satisfiability (MaxSAT) solver to find the most concise patches. 
Angelix \cite{Mechtaev2016} uses a lightweight repair constraint representation called ``angelic forest” to increase the scalability of DirectFix.

\subsection{Studies analyzing generated patches}

Recent studies have analyzed the patches generated by repair approaches from the literature.
The results of those studies show that generated patches may just overfit the available test cases, meaning that they will break untested but desired functionality. 
For example, Qui et al. \cite{Qi2015} find, using Kali system,  that the vast majority of patches produced by GenProg, RSRepair, and AE avoid bugs simply by functionality deletion. 
A subsequent study by Smith et al. \cite{smith2015cure} further confirms that the patches generated by  GenProg and RSRepair fail to generalize.

Due to the problematic of test overfitting, recent works by \cite{Liu2017IPC,Zu2017Test4Repair} propose to extend existing automated repair approach such as Nopol, ACS and jGenProg.
Those extended approaches generate new test inputs to enhance the test suites and use their behavior similarity to determine patch correctness.
For example, Liu  reported \cite{Liu2017IPC} that their approach, based on patch and test similarity analysis,  successfully prevented 56.3\% of the incorrect patches to be generated, without blocking any correct patches.
Yang et al. presented a framework named Opad (\underline{O}verfitted \underline{PA}tch \underline{D}etection) \cite{Yang2017BTC} to detect overfilled patches by enhancing existing test cases  using fuzz testing and employing two new test oracles. Opad  filters out 75.2\% (321/427) overfitted patches generated by GenProg/AE, Kali, and SPR.

\section{Threat to Validity}
\label{sec:threats}

\paragraph{Internal validity}
A threat to the validity of our results 
relates to whether our implementation of existing repair approaches are faithful to the original C implementation.
We have developed jGenProg, jKali and jMutRepair based on our deep analysis of the algorithms of GenProg, Kali and Mutation Repair, respectively, presented in different publications \cite{Weimer2009,Qi2015,debroy2010using}.
In the case of GenProg, we clearly see GenProg as two separate things: a repair approach  and a tool. 
We encode jGenProg based on the material presented in the GenProg publications. 
As the code of Astor is publicly available in the platform GitHub, researchers and developers can review and evaluate them, and if required to propose improvements if they spot inconsistencies with the original approach.

\paragraph{External validity}
We have evaluated the repair approaches built over Astor on 357 buggy program revisions, in five unique software systems, from the Defects4J benchmark \cite{JustJE2014}.
One threat is that the number of bugs may not be large enough nor not representative.
To mitigate this threat, we have executed repair approaches from Astor over other datasets of Java bugs (e.g., \cite{Ye2018Arxiv}), and we have found that Astor is also capable to repair bugs from an alternative dataset.

\paragraph{Summary} 
jGenProg has random components such as the selection of suspicious statements to modify. 
It is possible that different runs of jGenProg produce different patches. 
For this reason, we have executed each repair attempt with at least three different random seeds.
This is not meant to be a comprehensive solution to randomness. However, our goal is to validate our implementation, not the core idea, which was validated in the original publications.
Despite this small number of seeds, our experiments are computationally expensive due to the large number of bugs and the combinatorial explosion of repair approaches and configuration parameters. For instance, the experiment presented in Section  \ref{sec:deep2} consists of 19,949 trials spanning 2,616 days of computation time \cite{white2017dl}.
Moreover, as studied by different works \cite{Qi2015, smith2015cure,Le:2018:OSA,Yu2018,Xin:2017:ITP, defects4j-repair},  automated generated patches can suffer from  \emph{Overfitting}.
Those patches are correct with respect to the test suites used for validating them (they are \emph{test-suite adequate patched}), but yet incorrect. One of the reasons of accepting those patches is that the specification used by a repair approach (such as test-suites in the cases of test-suite based repair approaches) can be incomplete: a test suite does not include any input that triggers the unexpected -and incorrect- behaviour of a patch.
In our previous study, we manually analyzed patches generated by approaches built Astor, found that a portion of them overfit the evaluation test suites.
However, we believe that the overfitting problem affects to the repair approaches rather than the repair framework itself.
Astor provides  to developers extensions points for designing a new generation of repair approaches that aim at reducing the overfitting.

\section{Conclusion}
\label{sec:conclusion}

In this paper we presented Astor, a novel framework developed in Java that encodes the design space of generate-and-validate program repair approaches.
The framework contains the implementation of 6 repair approaches. It uniquely  provides extension points for facilitating research in the field.
The built-in repair approaches provided by Astor have already been used by researchers during the evaluation of their new repair approaches. Moreover, researchers have already implemented new components for Astor's extension points.
This paper presented an evaluation of the approaches provided by Astor, which repair 98 real bugs from the Defects4J dataset. 
We hope that Astor will facilitate the construction of new repair approaches and comparative evaluations in future research in automatic repair.
Astor is publicly available at \url{https://github.com/SpoonLabs/astor}.

\bibliographystyle{plain}

\bibliography{biblio-software-repair}

\end{document}